\documentclass[prl,floatfix,reprint,amsmath,amssymb,aps,longbibliography]{revtex4-2}

\usepackage{graphicx}
\usepackage{dcolumn}
\usepackage{bm}
\usepackage{chemformula}
\usepackage{indentfirst}
\usepackage[colorlinks=true, allcolors=blue]{hyperref}
\newcommand{\ii}{\mathrm{i}}


\begin{document}

\preprint{APS/123-QED}

\title{Hyperbolic exciton--plasmon polaritons in
\ch{MoS2}/\ch{MoOCl2} van der Waals heterostructures}

\author{Rajveer Fandan}
\email{rajveer.fandan@upm.es}
\affiliation{Instituto de Sistemas Optoelectrónicos y Microtecnología, Universidad Politécnica de Madrid, Av. Complutense 30, Madrid 28040, Spain}
\affiliation{Departamento de Ingeniería Electrónica, E.T.S.I de Telecomunicación, Universidad Politécnica de Madrid, Av. Complutense 30, Madrid 28040, Spain}

\author{Jorge Pedrós}
\email{j.pedros@upm.es}
\affiliation{Instituto de Sistemas Optoelectrónicos y Microtecnología, Universidad Politécnica de Madrid, Av. Complutense 30, Madrid 28040, Spain}
\affiliation{Departamento de Ingeniería Electrónica, E.T.S.I de Telecomunicación, Universidad Politécnica de Madrid, Av. Complutense 30, Madrid 28040, Spain}

\begin{abstract}
Polaritons formed by strong coupling between excitons and confined electromagnetic modes underpin emerging nanophotonic technologies, yet platforms based on isotropic metals or optical microcavities provide limited control over propagation direction. Here we predict that monolayer \ch{MoS2} on the in-plane hyperbolic conductor \ch{MoOCl2} supports hyperbolic exciton--plasmon polaritons governed by crystallographic direction, slab thickness, and hyperbolic mode order. Using an anisotropic transfer-matrix model coupled to a dissipative three-level Hamiltonian, we obtain wavevector-dependent anticrossings between the \ch{MoOCl2} plasmons and the spin--orbit--split A and B excitons of \ch{MoS2}, yielding lower, middle, and upper polariton branches with coupling energies $g_A=64.7~\mathrm{meV}$ and $g_B=61.2~\mathrm{meV}$. Rotating the in-plane wavevector tunes the plasmon--exciton detuning, while increasing the \ch{MoOCl2} thickness activates higher-order hyperbolic Fabry--P\'erot modes. Their coupling follows an approximate effective-mode-volume scaling. Strong coupling persists for excitonic linewidths up to $50~\mathrm{meV}$, with the normalised spectral-resolution ratio remaining above unity. These results establish \ch{MoS2}/\ch{MoOCl2} as a lithography-free platform for directional and multimode exciton--plasmon polariton engineering in the visible spectral range.
\end{abstract}

\maketitle

\section{Introduction}

The strong coupling of excitonic resonances in atomically thin
semiconductors with confined electromagnetic modes has emerged as a powerful platform for engineering hybrid light--matter
quasiparticles---polaritons---at deeply subwavelength
scales~\cite{mak2016photonics,liu2015strong,
dufferwiel2015exciton}.  Monolayer
transition-metal dichalcogenides (TMDs), and \ch{MoS2} in
particular, provide an ideal excitonic platform: the direct
band gap at the K valleys produces tightly bound A and B
excitons with large oscillator strengths and a spin--orbit
splitting of $\sim150$--$200$~meV~\cite{robert2018optical,mak2012control,luo2017opto}.  When these excitons are placed in
the near-field region of a metallic nanostructure, coupling to
localised or propagating surface plasmons can drive the system
from the Purcell-enhanced weak-coupling regime to the strong-coupling regime, where coherent energy exchange produces new hybrid
eigenstates~\cite{liu2016strong,wen2017room,cuadra2018observation,kim2022plasmon}.

Most reported TMD exciton--plasmon strong-coupling platforms
use Au or Ag, whose bulk optical response is isotropic.
Directionality must therefore be encoded in the fabricated
geometry---for example, through particle shape, grating period,
or gap size---and is fixed after fabrication.  Polariton
propagation and composition that depend directly on
crystallographic orientation require an intrinsically
anisotropic electromagnetic environment.

In-plane hyperbolic media provide such an environment~\cite{ma2024plane}.
The real parts of at least two principal dielectric-tensor
components have opposite signs, producing
open hyperbolic iso-frequency contours that channel
electromagnetic energy along preferred crystallographic
directions.  Among van der Waals conductors, \ch{MoOCl2} stands
out since its layered
oxyhalide structure gives rise to metallic conductivity along
one in-plane axis and insulating behaviour along the
orthogonal axis, yielding a natural in-plane hyperbolic window
across the visible spectrum~\cite{ruta2025good,gao2021robust}.
Low-loss hyperbolic plasmon polaritons have been observed in
exfoliated \ch{MoOCl2} flakes, and measurements of the full
dielectric tensor reveal a giant birefringence and a
visible-frequency epsilon-near-zero
crossing~\cite{li2026hyperbolic,venturi2024visible,
ermolaev2026giant,li2025broadband}.  Crucially, these
properties are intrinsic to the crystal and do not
require lithographic patterning, offering a scalable and
atomically precise alternative to metamaterial-based hyperbolic surfaces.

The \ch{MoOCl2} hyperbolic response ($\sim1.5$--$2.5$~eV)
overlaps the A and B excitons of monolayer \ch{MoS2}
($\sim1.9$--$2.1$~eV).  Stacking these materials can therefore
produce exciton--plasmon polaritons governed by the crystal
anisotropy of \ch{MoOCl2}.  Their near-field amplitude,
polarisation, and wavevector depend on the in-plane propagation direction.  A finite-thickness \ch{MoOCl2} slab also supports a ladder of hyperbolic Fabry--P\'{e}rot modes, providing thickness-controlled multimode coupling to the same excitonic transitions.

Here, we develop a theoretical framework for hyperbolic exciton--plasmon polaritons in \ch{MoS2}/\ch{MoOCl2} heterostructures. Combining an anisotropic transfer-matrix calculation with a non-Hermitian three-level model, we map the wavevector- and angle-dependent dispersion, quantify the modal composition, and evaluate the robustness of strong coupling against dissipation.

\section{Theoretical framework}

\subsection{Heterostructure geometry and electromagnetic model}

The system, sketched in Fig.~\ref{fig:1}a, consists of a monolayer
\ch{MoS2} placed on a \ch{MoOCl2} slab of
thickness $d$, which rests on a dielectric substrate with $\varepsilon_{sub} = 4$ corresponding to \ch{SiO2} substrate; the
superstrate is vacuum.  \ch{MoOCl2} is described by a
frequency-dependent diagonal dielectric tensor
$\boldsymbol{\varepsilon}(\omega)=
\mathrm{diag}[\varepsilon_x(\omega),\varepsilon_y(\omega),
\varepsilon_z(\omega)]$,
with each component parameterised by a Lorentz--Drude model
fitted to the experimentally determined optical
constants~\cite{ermolaev2026giant} (Supporting
Information, Section~S1). For a TM-polarised wave propagating
at angle $\theta$ with respect to the crystallographic $x$
axis, the in-plane response enters through the effective
permittivity
$\varepsilon_t(\omega,\theta)=\varepsilon_x\cos^{2}\!\theta
+\varepsilon_y\sin^{2}\!\theta$,
and the out-of-plane wavevector inside the slab is
\begin{equation}
k_{z}^{\mathrm{slab}}=
\sqrt{\varepsilon_t\,\omega^{2}/c^{2}
-(\varepsilon_t/\varepsilon_z)\,q^{2}},
\end{equation}
where $q$ is the in-plane wavevector magnitude. This directional
description is exact along the principal axes and is used as a
quasi-electrostatic approximation at arbitrary angles; off-axis
TE--TM mixing is neglected (Supporting Information, Section~S2). The
sign-changing anisotropy of $\varepsilon_t$ and $\varepsilon_z$
produces frequency windows where the large-$q$ contribution to $k_z$ becomes real, enabling propagating hyperbolic modes with deeply subwavelength in-plane wavevectors.

Monolayer \ch{MoS2} is treated as an infinitesimally thin
conducting sheet characterised by a wavevector-dependent surface
conductivity~\cite{vasilevskiy2015exciton}
\begin{equation}
\sigma(\omega,q)=
-\frac{\ii}{\omega}\sum_{n=A,B}
\frac{F_n}{E_n(q)-\hbar\omega-\ii\gamma_n/2},
\label{eq:sigma}
\end{equation}
where $E_n(q)=E_n+\hbar^{2}q^{2}/2M_n$ is the
wavevector-dependent exciton energy, $M_n$ is the exciton mass, $\gamma_n$ is its full width at
half maximum, and $F_n=4e^{2}v^{2}/\pi a_n^{2}$ encodes the
oscillator strength via the interband velocity $v$ and the exciton
Bohr radius $a_n$
(Supporting Information, Section~S3).

The electromagnetic response is obtained with the anisotropic
transfer-matrix model described in Supporting Information,
Section~S2. The \ch{MoS2} sheet
conductivity modifies the interface boundary
conditions~\cite{fandan2021exciton,fandan2018acoustically}. The hybrid modes are identified
through the optical loss function
\begin{equation}
L(\omega,\mathbf{q})=
\mathrm{Im}\bigl[r_{\mathrm{TM}}(\omega,\mathbf{q})\bigr],
\end{equation}
where $r_{\mathrm{TM}}$ is the TM-polarised reflection coefficient of
the complete multilayer stack.

\subsection{Non-Hermitian coupled-oscillator model}

The polariton spectrum is analysed quantitatively by
coupling the numerically extracted plasmon dispersion to two exciton states through a
dissipative three-level Hamiltonian (Supporting
Information, Section~S4),
\begin{equation}
H(q)=
\begin{pmatrix}
\tilde{E}_{\mathrm{pl}}(q) & g_{A} & g_{B}\\
g_{A} & \tilde{E}_{A} & 0\\
g_{B} & 0 & \tilde{E}_{B}
\end{pmatrix},
\label{eq:H3}
\end{equation}
where $\tilde{E}_{i}=E_{i}-\ii\gamma_{i}/2$ are the complex
self-energies for $i=\mathrm{pl},A,B$.  The
off-diagonal terms $g_{A}$ and $g_{B}$ are the coherent
coupling energies connecting the plasmon of \ch{MoOCl2} to the A and B
excitons of \ch{MoS2}, respectively; direct A--B coupling is
neglected because the A and B excitons originate from
spin--orbit--split valence bands with opposite spin character,
making the electric-dipole transition between them forbidden
to leading order~\cite{molina2013effect,wu2015exciton}.

The plasmon energy $E_{\mathrm{pl}}(q)$ and linewidth
$\gamma_{\mathrm{pl}}(q)$ are extracted from the peak position and
full width at half maximum of $L(\omega,q)$ computed for the
bare \ch{MoOCl2} slab. Smooth shape-preserving interpolation is used
between the extracted points. Diagonalisation of
Eq.~\eqref{eq:H3} at each $q$ yields three complex
eigenvalues
$\tilde{E}_{n}(q)=E_{n}(q)-\ii\Gamma_{n}(q)/2$,
defining the lower (LP), middle (MP), and upper (UP) polariton
branches. The corresponding right eigenvectors are normalised as
$\sum_i|C_i^{(n)}|^2=1$ and furnish the non-negative modal weights
$X_{i}^{(n)}=|C_{i}^{(n)}|^{2}$
($i=\mathrm{pl},A,B$) that quantify the plasmonic and
excitonic composition of each branch (Supporting
Information, Section~S5).

In the two-level limit (single exciton coupled to one plasmon),
the zero-detuning Rabi splitting reduces to (Supporting
Information, Section~S6)
\begin{equation}
\Omega_{R}=
2\sqrt{g^{2}-\left(\gamma_{\mathrm{pl}}-\gamma_{\mathrm{ex}}\right)^{2}/16}.
\label{eq:Rabi}
\end{equation}
Equation~\eqref{eq:Rabi} gives the dissipative eigenvalue splitting, whereas $S=4g/(\gamma_{\mathrm{pl}}+\gamma_{\mathrm{ex}})$ is used as a practical spectral-resolution criterion. Hereafter, the subscript on $\Omega$ and $S$ identifies the exciton under consideration.

\section{Results And Discussion}

\subsection{Anisotropic hyperbolic plasmon dispersion in
\ch{MoOCl2}}

\begin{figure*}[!t]
\centering
\includegraphics[width=\textwidth]{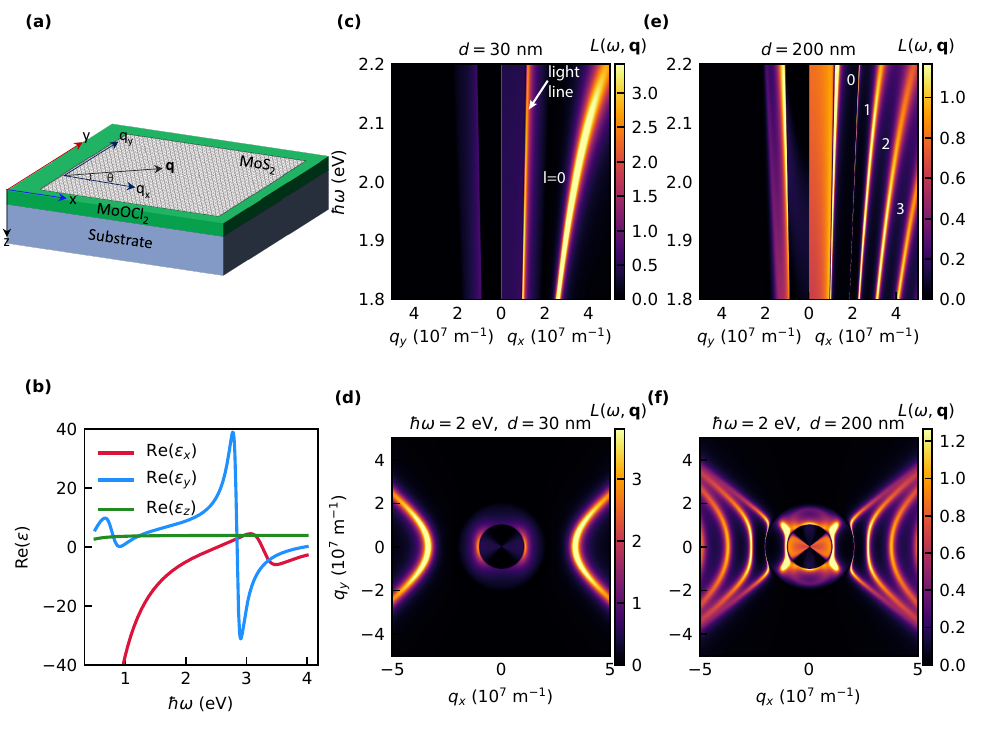}
\caption{\textbf{Anisotropic hyperbolic plasmon modes in \ch{MoOCl2}.}
\textbf{a},~Schematic of the heterostructure: monolayer \ch{MoS2}
(exciton sheet) on an anisotropic \ch{MoOCl2} slab of thickness $d$
supported by a dielectric substrate ($\varepsilon_{\mathrm{sub}}=4$).  The
in-plane wavevector $\mathbf{q}=(q_x,q_y)$ is parameterised by its
magnitude $q$ and the azimuthal angle $\theta$ measured from the
metallic $x$ axis.
\textbf{b},~Real parts of the principal dielectric-tensor components
of \ch{MoOCl2} from the Lorentz--Drude model.
\textbf{c},~Wavevector-resolved loss function
$L(\omega,\mathbf{q})$ for $d=30~\mathrm{nm}$.  The left
(right) half shows propagation along $q_y$ ($q_x$).  The asymmetric dispersion
directly reflects the sign-changing in-plane anisotropy.
\textbf{d},~Iso-frequency loss map at $\hbar\omega=2.0~\mathrm{eV}$ for
$d=30~\mathrm{nm}$, showing the open hyperbolic contour of the fundamental
mode.
\textbf{e},~As in \textbf{c} but for $d=200~\mathrm{nm}$.  The increased
optical path length quantises the response into a ladder of
higher-order hyperbolic Fabry--P\'{e}rot branches.
\textbf{f},~Iso-frequency map at $\hbar\omega=2.0~\mathrm{eV}$ for $d=200~\mathrm{nm}$; the
concentric hyperbolic contours correspond to successive mode orders.}
\label{fig:1}
\end{figure*}

We first characterise the intrinsic electromagnetic modes of
the bare \ch{MoOCl2} slab.  Figure~\ref{fig:1}b shows the calculated
real parts of the three principal dielectric components.  In
the spectral window $1.5$--$2.5$~eV,
$\mathrm{Re}(\varepsilon_x)$ is large and negative while
$\mathrm{Re}(\varepsilon_y)$ remains positive, establishing an
in-plane type-II hyperbolicity (the metallic axis is $x$, the
dielectric axis is $y$).  This sign-changing anisotropy is the
microscopic origin of the directional plasmon confinement
discussed below.

Figure~\ref{fig:1}c presents the wavevector-resolved loss function for a
$d=30~\mathrm{nm}$ slab, displayed as a split panel with the left and
right halves corresponding to propagation along $y$ and $x$,
respectively.  A single well-defined plasmon branch is
visible, but its dispersion is strikingly asymmetric: along the
metallic $x$ axis the mode disperses to high wavevector with
strong confinement, whereas along $y$ it is cut off at
smaller wavevectors.  Physically, this asymmetry arises because
$\varepsilon_x<0$ supports a metallic screening response that
sustains large-$q$ plasmonic oscillations, while
$\varepsilon_y>0$ provides only dielectric restoring forces.
The iso-frequency contour for $d=30~\mathrm{nm}$ at $\hbar\omega=2.0~\mathrm{eV}$
[Fig.~\ref{fig:1}d] confirms the open hyperbolic topology of the
wavevector-space surface, in stark contrast to the closed
circular contours of isotropic metals.

Increasing the slab thickness to $d=200~\mathrm{nm}$ [Fig.~\ref{fig:1}e]
produces a discrete set of higher-order hyperbolic modes,
analogous to Fabry--P\'{e}rot standing waves in an
anisotropic dielectric cavity.  These modes arise because the
increased optical path length allows multiple half-wavelengths
to fit within the slab along the out-of-plane direction.  Each
mode order $l$ has a distinct transverse field profile
$\sim\!\cos(l\pi z/d)$ or $\sim\!\sin(l\pi z/d)$ and a different
degree of confinement.  The fraction of modal energy stored in
the slab therefore varies with $l$ and controls the coupling
discussed below.  The corresponding iso-frequency contour at
$2.0~\mathrm{eV}$ [Fig.~\ref{fig:1}f] reveals multiple concentric hyperbolic
branches, each associated with a different mode order.

\subsection{Formation of three-branch exciton--plasmon
polaritons}

\begin{figure*}[!t]
\centering
\includegraphics[width=\textwidth]{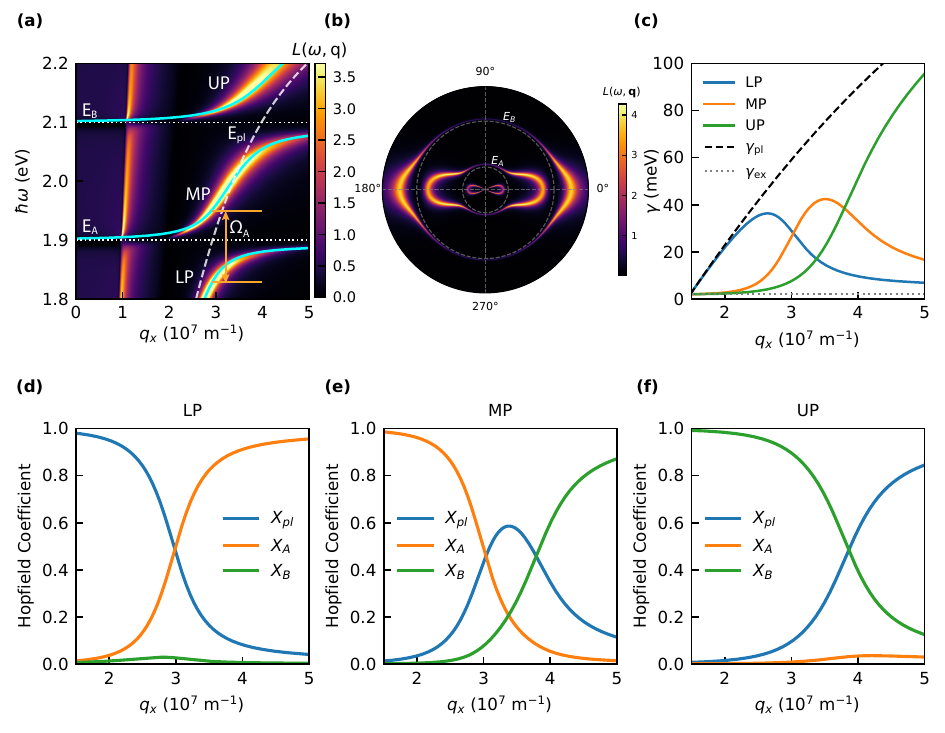}
\caption{\textbf{Wavevector-space polariton dispersion, anisotropic response, and modal composition of exciton--plasmon polaritons.}
\textbf{a},~Calculated loss function $L(\omega,q_x)$ for a
30-nm-thick \ch{MoS2}/\ch{MoOCl2} heterostructure. The colour map
shows the optical loss intensity, while the cyan curves are the
polariton eigenenergies obtained from the non-Hermitian
coupled-oscillator model. The dashed white curve indicates the bare
plasmon dispersion and the horizontal dotted lines mark the uncoupled
A- and B-exciton resonances. The avoided crossings reveal the
formation of lower (LP), middle (MP), and upper (UP) polariton branches.
\textbf{b},~Angular dependence of the loss function
$L(\omega,\mathbf{q})$ at fixed wavevector magnitude
$q=3.97\times10^{7}~\mathrm{m}^{-1}$, shown in polar coordinates.
The anisotropic response reflects the directional hyperbolic plasmon
dispersion of \ch{MoOCl2}.
\textbf{c},~Calculated linewidths of the LP, MP, and UP branches
obtained from the imaginary parts of the complex eigenenergies.
The dashed black curve shows the bare plasmon linewidth
$\gamma_{\mathrm{pl}}$, while the dotted gray line indicates the
exciton linewidth $\gamma_{\mathrm{ex}}$. The wavevector-dependent
linewidth redistribution reflects the transfer of plasmonic and
excitonic dissipation among the hybrid polariton modes.
\textbf{d--f},~Modal weights of the LP, MP, and UP branches,
respectively. The coefficients $X_{\mathrm{pl}}$, $X_A$, and $X_B$
represent the plasmonic, A-excitonic, and B-excitonic components,
showing the exchange of modal character across the anticrossings.}
\label{fig:2}
\end{figure*}

Having established the bare plasmon landscape, we now include
the monolayer \ch{MoS2} conductivity in the transfer-matrix
calculation.  Figure~\ref{fig:2}a shows the resulting loss function
for $d=30~\mathrm{nm}$ with propagation along the $x$ axis
($\theta=0$).  Two pronounced anticrossings are visible where
the plasmon branch intersects the A ($E_A=1.90~\mathrm{eV}$) and B
($E_B=2.10~\mathrm{eV}$) exciton energies.

Diagonalisation of the three-level Hamiltonian
[Eq.~\eqref{eq:H3}] with coupling strengths
$g_{A}=64.7~\mathrm{meV}$ and $g_{B}=61.2~\mathrm{meV}$
yields polariton dispersions (cyan curves) that agree with the
full electromagnetic loss spectrum.  These coupling energies
satisfy the strong-coupling criterion of
Eq.~\eqref{eq:Rabi}: using the
numerically extracted plasmon linewidth
$\gamma_{\mathrm{pl}}\approx 80~\mathrm{meV}$ and excitonic linewidths
$\gamma_{\mathrm{ex}}=\gamma_{A}=\gamma_{B}=2~\mathrm{meV}$, the
dissipation-corrected
Rabi splittings are $\Omega_{A}\approx 123~\mathrm{meV}$ and
$\Omega_{B}\approx 116~\mathrm{meV}$. The corresponding spectral-resolution
ratios are $S_A\approx3.16$ and $S_B\approx2.99$, placing both
anticrossings in the strong-coupling regime (Supporting
Information, Section~S5).

\paragraph{Directional control of hybridisation.}

The polar representation of the loss function at the fixed wavevector
$q=3.97\times10^{7}~\mathrm{m}^{-1}$, corresponding to the B-exciton
anticrossing in Fig.~\ref{fig:2}a, reveals a strongly anisotropic
polariton spectrum [Fig.~\ref{fig:2}b]. Because
the plasmon energy depends on the propagation angle $\theta$
through $\varepsilon_t(\omega,\theta)$, rotating the in-plane
wavevector continuously tunes the plasmon--exciton detuning
without modifying any material or geometric parameter.
Along the metallic axis ($\theta=0$), the plasmon energy at
this $q$ lies between the A and B excitons, producing
two well-resolved anticrossings.  Rotating toward the
dielectric axis ($\theta=90^{\circ}$), the plasmon blue-shifts
and the system moves away from resonance, weakening the
hybridisation.  In an isotropic plasmonic medium, comparable
directional tuning would require changing the geometry.

\paragraph{Polariton composition and dissipation.}

The modal weights [Fig.~\ref{fig:2}d--f] quantify the
continuous redistribution of oscillator character across the
anticrossing regions (Supporting
Information, Section~S5).  At small $q$, the LP is predominantly
plasmonic; as $q$ increases past the first anticrossing, it
acquires a dominant A-excitonic character.  The MP undergoes the
reverse transition and subsequently hybridises with the B
exciton at the second anticrossing.  The UP evolves from
B-exciton-like at intermediate $q$ toward a mixed
plasmon--exciton state at large $q$.

The polariton linewidths [Fig.~\ref{fig:2}c] track the mode
composition: branches with a larger plasmonic fraction inherit
the broader plasmon damping, while exciton-dominated segments
are spectrally narrower.  Importantly, even in the maximally
mixed states at the anticrossing regions the linewidths remain smaller than
the mode splitting, confirming that the polaritons are
well-resolved hybrid eigenstates rather than overlapping
spectral features.

\subsection{Multimode coupling controlled by slab thickness
and hyperbolic mode order}
A key advantage of a hyperbolic polaritonic platform over a conventional
isotropic interface is that a single slab can support a ladder
of guided modes rather than a single surface mode. In the spectral
range of the A and B excitons, \ch{MoOCl2} exhibits type-II hyperbolic
dispersion, enabling the formation of volume-confined hyperbolic modes
with large in-plane wavevectors. At the A-exciton energy
($\hbar\omega=1.9~\mathrm{eV}$), for $\theta=0$ the dielectric response obtained from our
dispersion model is
$\varepsilon_t\equiv\varepsilon_x\simeq-6.3+0.4\ii$ and
$\varepsilon_z\simeq3.9$. In the high-wavevector limit, the
out-of-plane wavevector is approximately

\begin{equation}
k_z\simeq q\sqrt{\frac{|\varepsilon_t|}{\varepsilon_z}},
\end{equation}

which is real, indicating oscillatory field confinement
across the slab thickness rather than evanescent decay. The slab
therefore acts as a Fabry--P\'erot resonator for hyperbolic modes, where
the allowed wavevectors are determined by the phase accumulation across the
thickness,

\begin{equation}
q_l d
\sqrt{\frac{|\varepsilon_t(\omega)|}{\varepsilon_z(\omega)}}
=
l\pi+\phi_{\mathrm{air}}+\phi_{\mathrm{sub}},
\quad
\tan\phi_j=
\frac{\varepsilon_j}
{\sqrt{\varepsilon_z|\varepsilon_t|}},
\label{eq:quant}
\end{equation}

where $q_l$ is the in-plane wavevector of the $l$th hyperbolic mode, and $\phi_{\mathrm{air}}$ and $\phi_{\mathrm{sub}}$
are the reflection phase shifts at the upper and lower interfaces,
respectively. Here, $\varepsilon_j$ denotes the relative permittivity
of the surrounding dielectric medium, with
$j\in\{\mathrm{air},\mathrm{sub}\}$. The integer
$l=0,1,2,\dots$ labels the transverse mode order. Consequently,
changing the slab thickness or mode order provides direct control over
the in-plane wavevector and confinement of the hyperbolic modes.
Increasing the mode order generally increases the wavevector and
confinement, whereas increasing the slab thickness reduces the
wavevector of a given mode.

\begin{figure*}[!t]
\centering
\includegraphics[width=\textwidth]{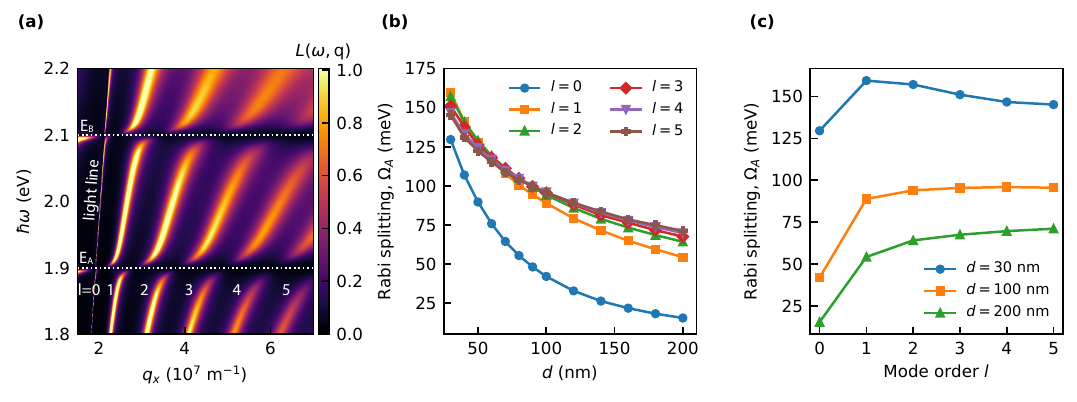}
\caption{\textbf{Multimode exciton--plasmon coupling controlled
by the thickness and hyperbolic mode order.}
\textbf{a},~Loss function $L(\omega,q)$ of the
\ch{MoS2}/\ch{MoOCl2} heterostructure for $d=200$~nm. Multiple
volume-confined hyperbolic Fabry--P\'erot branches, quantized
according to Eq.~\eqref{eq:quant}, cross the A and B exciton
energies (white dotted lines) and produce independent
anticrossings at distinct in-plane wavevectors $q_l$.
\textbf{b},~A-exciton Rabi splitting $\Omega_A$ versus
\ch{MoOCl2} thickness for mode orders $l=0$--$5$. For
well-confined modes, $\Omega_A\propto d^{-1/2}$, consistent
with the expected scaling for volume-confined hyperbolic modes.
The much steeper decay of the $l=0$ (and $l=1$) branches
reflects the loss of confinement as the lowest-order
hyperbolic modes approach the substrate light line.
\textbf{c},~$\Omega_A$ versus mode order $l$ for
$d=30$, $100$, and $200$~nm. $\Omega_A$ increases with
$l$ and saturates once the modes become well confined
($\Gamma_l\rightarrow1$). The saturation shifts to higher
mode order for thicker slabs because larger in-plane wavevectors
are required to achieve strong confinement.}
\label{fig:3}
\end{figure*}

As a consequence, several hyperbolic modes can cross the
excitonic resonance, with each mode coupling to the excitonic
centre-of-mass state at its corresponding wavevector $q_l$. Figure~\ref{fig:3}a
illustrates this multimode coupling for a $200$-nm-thick
\ch{MoOCl2} slab. Multiple hyperbolic branches intersect the A and B
exciton energies and generate separate anticrossings at different values
of $q_l$.

For a single excitonic transition, the coupling energy scales as

\begin{equation}
g_l \propto
\frac{1}{\sqrt{V_{\mathrm{eff}}^{(l)}}},
\end{equation}

where $V_{\mathrm{eff}}^{(l)}$ is the effective mode volume sampled by
the exciton~\cite{vahala2003optical,marquier2017revisiting}. For modes
extended across the slab thickness,
\begin{equation}
V_{\mathrm{eff}}^{(l)}
\propto
\frac{d}{\Gamma_l},
\end{equation}
where $\Gamma_l$ is the confinement factor, which accounts for leakage into the surrounding media.

In the low-loss approximation, the Rabi splitting extracted at the
A-exciton resonance wavevector follows
\begin{equation}
\Omega_A^{(l)}
\simeq
2\sqrt{\frac{\mathcal{C}\Gamma_l}{d}},
\label{eq:scaling}
\end{equation}

where $\mathcal{C}$ contains the excitonic oscillator strength and other
material parameters. $\Gamma_l$ can be estimated by approximating the modal fields as exponentially decaying outside the hyperbolic slab (Supporting
Information, Section~S7),

\begin{equation}
\Gamma_l=
\frac{1}{
1+\dfrac{1}{2\kappa_l^{\mathrm{air}}d}
+\dfrac{1}{2\kappa_l^{\mathrm{sub}}d}},
\label{eq:gamma}
\end{equation}

where

\begin{equation}
\kappa_l^{j}
=
\sqrt{q_l^2-\varepsilon_j\frac{\omega^2}{c^2}}
\end{equation}

is the decay constant of the electromagnetic field in the surrounding
medium $j\in\{\mathrm{air},\mathrm{sub}\}$. When $q_l$ greatly exceeds
the corresponding light-line wavevectors, the fields decay rapidly
outside the slab and $\kappa_l d\gg1$; Eq.~\eqref{eq:gamma} then gives
$\Gamma_l\rightarrow1$. In this regime the
electromagnetic field is primarily localised within the hyperbolic
medium, and the coupling approaches the ideal volume-confined scaling

\begin{equation}
\Omega_A\propto d^{-1/2}.
\end{equation}

Conversely, when a mode approaches a light line, the
evanescent decay length increases, the electromagnetic field extends
further into the surrounding media, and $\Gamma_l$
decreases, reducing the exciton--field interaction.

The extracted splittings in Fig.~\ref{fig:3}b follow this picture.
Higher-order modes (especially $l\ge1$) remain strongly confined over most of the
investigated thickness range and approximately follow the expected
$d^{-1/2}$ scaling. The lowest-order branch deviates from this behaviour
because its wavevector decreases most rapidly with increasing slab
thickness. For sufficiently large $d$, the $l=0$ mode approaches the
substrate light line and becomes weakly confined, causing a reduction of
$\Gamma_0$ and a stronger suppression of $\Omega_A$.

The dependence on mode order is summarised in Fig.~\ref{fig:3}c.
At a fixed thickness, increasing $l$ generally increases $q_l$ and
improves confinement, causing the Rabi splitting to increase until the
confinement-limited regime is reached. Once the mode is sufficiently far away from the light line, $\Gamma_l$ approaches unity and the
coupling approaches the value set by the excitonic oscillator strength
and slab-confined mode volume. Thus, confinement rather than mode order
itself controls the interaction.

\begin{figure}[!t]
    \centering
    \includegraphics[width=\columnwidth]{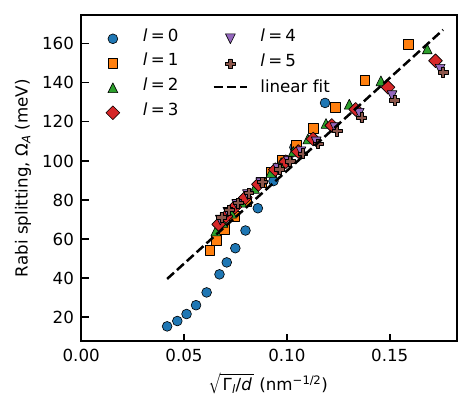}
    \caption{\textbf{Scaling of the multimode Rabi splitting with the effective confinement parameter $\sqrt{\Gamma_l/d}$}.
The extracted Rabi splittings $\Omega_A$ for the hyperbolic modes $l=0$--$5$ are plotted as a function of $\sqrt{\Gamma_l/d}$, where $\Gamma_l$ is the effective confinement factor describing the degree of electromagnetic confinement within the hyperbolic slab.
The dashed line shows the physically constrained scaling
$\Omega_A=(948\,\mathrm{meV\,nm^{1/2}})
\sqrt{\Gamma_l/d}$ obtained from a least-squares fit forced through the
origin ($R^2=0.902$).
Most higher-order modes follow this confinement scaling; the largest
deviation occurs for the $l=0$ mode at large thickness, where proximity
to the substrate light line increases leakage.}
    \label{fig:4}
\end{figure}

This scaling is further demonstrated in Fig.~\ref{fig:4}, where all
calculated Rabi splittings are plotted against
$\sqrt{\Gamma_l/d}$. The approximately linear dependence confirms that
the thickness and mode-order dependence of the exciton--hyperbolic mode
coupling is primarily governed by the effective mode volume. Since the
coupling must vanish in the limit of infinite mode volume
(Eq.~\eqref{eq:scaling}), we fit the data using the physically
constrained relation

\begin{equation}
\Omega_A =
948~\mathrm{meV\,nm^{1/2}}
\sqrt{\frac{\Gamma_l}{d}},
\end{equation}

where $d$ is expressed in nm. The constrained fit yields a coefficient
of determination of $R^2=0.902$, demonstrating that the simple
effective-mode-volume model captures the dominant dependence of the
exciton--hyperbolic mode coupling on slab thickness and modal
confinement. Most higher-order modes follow this approximate scaling,
while the largest deviations originate from the $l=0$ branch
at large thicknesses, where proximity to the substrate light line
enhances the electromagnetic leakage into the surrounding media and
causes the simplified confinement factor $\Gamma_l$ to underestimate the
actual mode-volume correction. Thickness therefore controls the overall
mode volume, while mode order controls the confinement factor
$\Gamma_l$.

\begin{figure*}[!t]
\centering
\includegraphics[width=\textwidth]{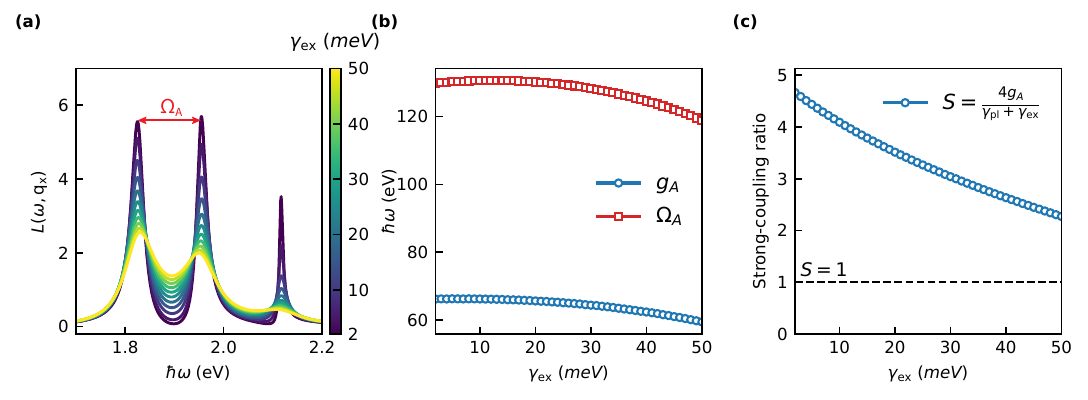}
\caption{\textbf{Robustness of strong coupling against
excitonic dissipation.}
\textbf{a},~Loss spectra at the anticrossing wavevector
$q_x=2.91\times10^{7}~\mathrm{m}^{-1}$ for exciton linewidths
ranging from $2$ to $50~\mathrm{meV}$.  The two polariton
peaks remain well resolved throughout.
\textbf{b},~Extracted Rabi splitting $\Omega_A$ and fitted effective coupling $g$ as functions
of the exciton linewidth $\gamma_{\mathrm{ex}}$.  Although both quantities decrease gradually with increasing excitonic
dissipation, the coupling strength remains sufficiently large
to sustain the strong-coupling regime.
\textbf{c},~Spectral-resolution
ratios
$S=4g/(\gamma_{\mathrm{pl}}+\gamma_{\mathrm{ex}})$ versus
$\gamma_{\mathrm{ex}}$.  The horizontal dashed line ($S=1$)
indicates the conventional threshold for spectrally resolved
strong coupling.  The system remains above this threshold over
the investigated linewidth range, demonstrating the robustness
of exciton--plasmon strong coupling against substantial
excitonic dissipation.}
\label{fig:5}
\end{figure*}

\subsection{Robustness of strong coupling against excitonic
dissipation}

The practical viability of any polariton platform depends on
whether the strong-coupling condition survives under realistic
material losses.  In monolayer TMDs, the excitonic linewidth is
highly sensitive to temperature, defect density, substrate
disorder, and dielectric environment, and can range from
${\sim}2$~meV at cryogenic temperatures to
${\sim}40$--$50$~meV at room
temperature~\cite{cadiz2017excitonic,selig2016excitonic}.
We therefore vary $\gamma_{\mathrm{ex}}$ from 2 to $50~\mathrm{meV}$
while holding the material and geometrical parameters fixed.

Figure~\ref{fig:5}a shows that the LP and MP peaks persist across
the full linewidth range.  They broaden and move closer but
remain spectrally resolved even at
$\gamma_{\mathrm{ex}}=50$~meV. The spectral peak separation
[Fig.~\ref{fig:5}b] decreases from $\sim125$ to
$\sim105$~meV, a reduction of $\sim16\%$ despite
a 25-fold increase in excitonic damping.  The fitted effective
coupling $g$ decreases moderately, from
approximately 70 to 60~meV, while remaining well above the
strong-coupling threshold. Because the bare Hamiltonian coupling is
fixed, the fitted $g$ is an effective spectral parameter rather than a
change in the microscopic interaction. Excitonic dissipation reduces
spectral resolvability but does not eliminate the mode splitting.

To quantify the persistence of spectrally resolved strong-coupling regime, we plot spectral-resolution
ratio $S$ in Fig.~\ref{fig:5}c. The horizontal dashed line at $S=1$ marks the
conventional criterion for spectrally resolved strong
coupling, corresponding to
$g=(\gamma_{\mathrm{pl}}+\gamma_{\mathrm{ex}})/4$.  Although $S$
decreases monotonically with increasing excitonic linewidth, it
remains above unity throughout the investigated range. The confined
hyperbolic mode therefore maintains spectrally resolved coupling under
substantial excitonic broadening. Thus, \ch{MoS2}/\ch{MoOCl2} polaritons can remain strongly coupled for linewidths representative of room-temperature monolayer TMDs.

The \ch{MoS2}/\ch{MoOCl2} heterostructure combines directional
polariton propagation with thickness-dependent multimode strong
coupling.

\paragraph{Directional polaritonics without lithography.}
In isotropic metals, directionality generally requires patterned
gratings, antennas, or waveguides. The intrinsic anisotropy of
\ch{MoOCl2} instead provides directional dispersion without
nanofabrication, so crystal orientation controls propagation,
wavevector matching, and light--matter interaction.

\paragraph{Multimode coupling as a design resource.}
The hyperbolic Fabry--P\'{e}rot ladder provides modes with distinct
wavevectors, field profiles, linewidths, and group velocities. Their
coupling to the same excitonic transition offers routes to interbranch
scattering, multimode dynamics, and multi-resonant sensing. This
multimode response is further enabled by the asymmetric
\ch{MoS2}/\ch{MoOCl2} heterostructure, which permits coupling to both
even- and odd-parity hyperbolic modes. By contrast, in a symmetric air/\ch{MoOCl2}/\ch{MoS2}/\ch{MoOCl2}/air structure, symmetry allows coupling only to the even-parity modes, consistent with previous theoretical work~\cite{eini2025strongly} (Supporting
Information, Section~S8).

\paragraph{Towards experimental realisation.}
Both \ch{MoS2} and \ch{MoOCl2} are exfoliable van der Waals
crystals~\cite{liu2015strong,ermolaev2026giant}, and the predicted
$\sim100$--$160~\mathrm{meV}$ splittings are accessible to current
nano-optical spectroscopy. Momentum-resolved electron-energy-loss
spectroscopy (EELS)~\cite{venturi2024visible} and scattering-type scanning near-field optical microscopy (s-SNOM) can probe large-wavevector modes~\cite{ermolaev2026giant} (Supporting
Information, Section~S8). Surface acoustic waves
offer a complementary coupling route: their periodic modulation acts
as a dynamic diffraction grating that supplies the additional in-plane
wavevector required to access high-$q$
polaritons~\cite{fandan2021exciton,ruppert2010surface}.

Strong coupling remains spectrally resolved up to
$\gamma_{\mathrm{ex}}=50~\mathrm{meV}$. Cryogenic operation would
further improve spectral visibility and propagation
length~\cite{ni2018fundamental}. The principal experimental challenge is
the deterministic alignment of the two crystals, because their relative
orientation fixes the azimuthal angle $\theta$.

\paragraph{Limitations and outlook.}
Our treatment assumes a spatially local excitonic response
characterised by Eq.~\eqref{eq:sigma}. At very large wavevectors
($q \gtrsim 1/a_{\mathrm{n}}$), nonlocal corrections to both the excitonic
conductivity and the plasmonic response may modify the polariton
dispersion in the extreme-confinement regime. The present model
treats the \ch{MoOCl2} dielectric response as a local bulk quantity;
for slabs thinner than approximately $5~\mathrm{nm}$, quantum confinement and
surface effects may modify the dielectric tensor and shift the
hyperbolic spectral window. Electronic correlations in \ch{MoOCl2},
suggested by transport measurements~\cite{ruta2025good}, are also not
captured by the Lorentz--Drude model and may affect plasmon damping and
spectral weight.

Extensions of this work could include twist-angle control and moir\'e
excitons~\cite{jin2019observation}, coupling to dark or interlayer
excitons~\cite{zhou2017probing,rivera2015observation,alexeev2019resonantly},
and nonlinear polariton phenomena~\cite{carusotto2013quantum}. The theoretical framework reported here is applicable to other anisotropic van der Waals conductors
coupled to atomically thin excitonic semiconductors.

\section{Conclusion}
We predict hyperbolic exciton--plasmon polaritons in a
\ch{MoS2}/\ch{MoOCl2} van der Waals heterostructures. Coupling one
hyperbolic plasmon branch to the spin--orbit--split A and B excitons
produces lower, middle, and upper polariton branches with
$g_A=64.7~\mathrm{meV}$ and $g_B=61.2~\mathrm{meV}$. Crystal
orientation controls the directional dispersion and detuning, whereas
slab thickness activates higher-order Fabry--P\'{e}rot modes. Their
splittings approximately scale as $\sqrt{\Gamma_l/d}$, and the
normalised ratio $S=4g/(\gamma_{\mathrm{pl}}+\gamma_{\mathrm{ex}})$
remains above unity for excitonic linewidths up to
$50~\mathrm{meV}$.

Anisotropic van der Waals conductors therefore provide a
lithography-free platform for directional and multimode
exciton--plasmon polaritonics. The predicted modes are accessible to
near-field, electron-beam, and surface-acoustic-wave-assisted
techniques, with opportunities for engineering the photonic density of
states, anisotropic Purcell enhancement, and mode-selective
light--matter interactions~\cite{poddubny2013hyperbolic}.

\vspace{\baselineskip}

\noindent\textbf{Data availability} The data supporting the findings of this study are available from the corresponding author upon reasonable request.

\vspace{\baselineskip}

\noindent\textbf{Competing interests} The authors declare no competing interests.

\vspace{\baselineskip}

\noindent\textbf{Acknowledgments}
This work has received funding from the Spanish Ministry of Science and Innovation (MICINN) through project 2D-SAWNICS (PID2020-120433GB-I00), from the Spanish Ministry of Science, Innovation, and Universities (MICIU) through project SE2D (PID2024-159409OB-I00), and from the Comunidad de Madrid, the Recovery, Transformation and Resilience Plan of the Spanish Government, and the European Union--NextGenerationEU (PRTR-C17.I1) through projects MAD2D-CM-UPM1 (Complementary Plan for Advanced Materials) and MADQuantum-CM (Complementary Plan for Quantum Communications).

\bibliography{reference}

\clearpage
\thispagestyle{empty}

\begin{center}
    \includegraphics[width=8.25cm,height=4.45cm]{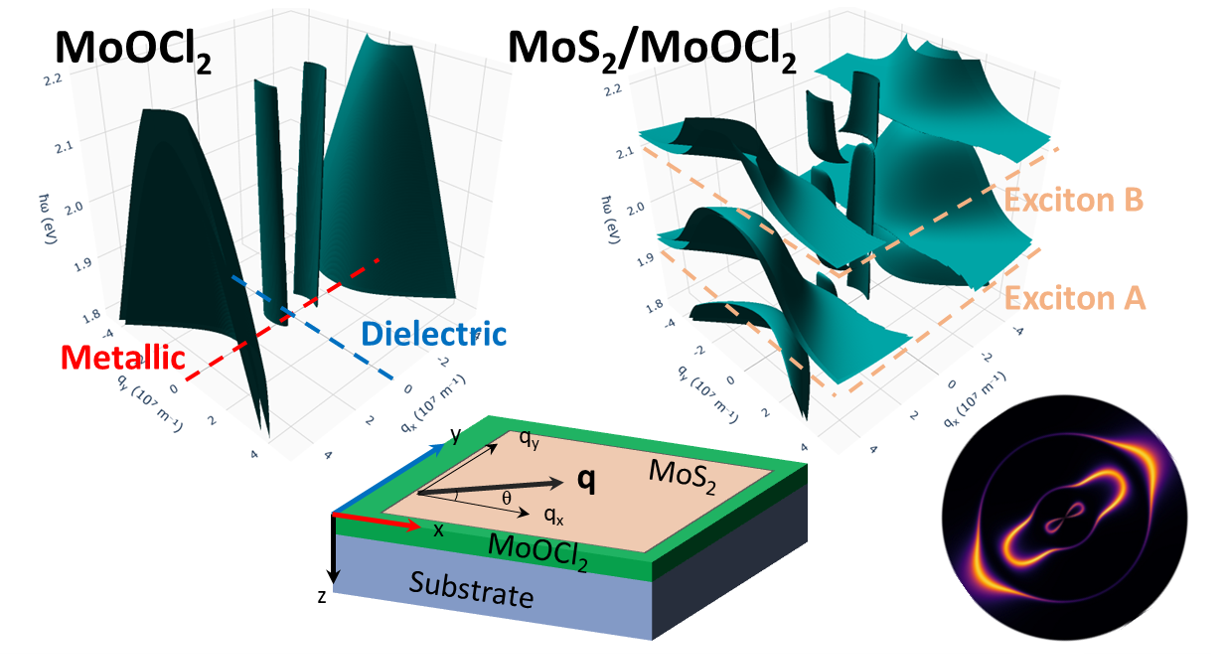}
\end{center}

\end{document}